# Mesure des risques de marche et de souscription vie en situation d'information incomplete pour un portefeuille de prevoyance

Jean-Paul Felix*                                   Frédéric Planchet*

BNP Paribas Assurance$^\alpha$
Université de Lyon - Université Claude Bernard Lyon 1
ISFA – Actuarial School $^\gamma$
WINTER & Associés $^\lambda$

## Résumé

Dans le cadre des nouvelles normes d'*Embedded Value*, les normes MCEV, les derniers principes publiés en juin 2008 abordent la question de la mesure des risques de marché et de souscription par l'intermédiaire de modèles stochastiques de projection et de valorisation.

Les modèles stochastiques étant fortement consommateurs en terme de données, la question qui se pose est le traitement de portefeuilles d'assurance uniquement disponibles en données agrégées ou de portefeuilles en situation d'information incomplète

L'objet de cet article est de proposer une modélisation pragmatique de ces risques attachés aux garanties décès des produits de prévoyance individuelle dans ces situations.

**MOTS-CLEFS :** Embedded Value, coût des risques résiduels non couvrables (CRNHR), options et garanties financières (TVFOG), prévoyance individuelle

## Abstract

In the framework of *Embedded Value* new standards, namely the MCEV norms, the latest principles published in June 2008 address the issue of market and underwriting risks measurement by using stochastic models of projection and valorization.

Knowing that stochastic models  particularly data-consuming, the question which can arise is the treatment of insurance portfolios only available in aggregate data or portfolios in situation of incomplete information.

The aim of this article is to propose a pragmatic modeling of these risks tied up with death covers of individual protection products in these situations.

* Contact : jean-paul.f.felix@bnpparibas.com, fplanchet@winter-associes.fr
$^\alpha$ BNP PARIBAS Assurance – 4 rue des frères Caudron – 92858 Rueil Malmaison Cedex
$^\gamma$ Institut de Science Financière et d'Assurances (ISFA) - 50 avenue Tony Garnier - 69366 Lyon Cedex 07 – France.
$^\lambda$ WINTER & Associés – 43-47 avenue de la Grande Armée - 75116 Paris et 18 avenue Félix Faure - 69007 Lyon – France.

## 1. Introduction

Dans le cadre de l'élaboration des nouvelles normes d'Embedded Value applicables au secteur de l'assurance, les normes MCEV[1], des discussions sont menées entre les directeurs financiers des principales sociétés d'assurance européennes au sein du CFO Forum.

L'objectif de ce groupe de discussion est de progresser dans l'harmonisation des normes comptables et d'information financière. Ainsi, les récents développements ont montré une convergence des assureurs en Europe pour évaluer leurs portefeuilles selon une approche *Market Consistent*. Les derniers principes publiés (cf. CFO Forum [2008]) abordant notamment les règles en matière de calcul et de communication de la MCEV constitueront, à partir du 31 décembre 2009, l'unique référence pour les Embedded Value calculées avec les principes du CFO Forum.

L'Embedded Value est une mesure de valeur composée de deux parties : le patrimoine accumulé au cours du temps et la valeur des contrats en portefeuille obtenue par projection en run off des flux de résultats projetés. Le calcul des éléments constituant cette valeur implique le recours de techniques actuarielles faisant appel aux approches stochastiques. Parmi les trois composantes de la MCEV (le capital requis, l'excédent de capital et la valeur du portefeuille – appelée VIF -), l'élément qui a le plus évolué est la valeur du portefeuille qui correspond à la valeur actuelle des profits futurs diminuée de la valeur temps des options et garanties, des coûts frictionnels et des risques résiduels non couvrables.

Pour le calcul de la valeur temps des options et garanties financières, les modèles stochastiques utilisés doivent être calibrés à partir des données de marché observables les plus récentes. De plus, les hypothèses de volatilité doivent correspondre dans la mesure du possible aux volatilités implicites déduites des prix des instruments dérivés échangés sur le marché. Cette valeur, ajoutée à la valeur intrinsèque des options et garanties, reflète le prix actuel d'une couverture acquise sur le marché, sans que le risque de crédit propre à l'entité soit intégré dans la valorisation.

Le coût des risques résiduels non couvrables correspond à l'impact sur la valeur de l'incertitude sur les résultats à venir relative aux risques qui ne peuvent pas être couverts sur les marchés financiers. Dans le cadre de cet article, nous nous attarderons sur le risque de souscription vie.

Eléments centraux de la matérialisation des modèles internes, les systèmes d'information et les données à traiter sont au centre des préoccupations des assureurs. Dans les réflexions menées actuellement, les données sont supposées être stables dans le temps, avec des historiques normalisés, et d'une grande finesse. Cette grande finesse entraîne toutefois un certain nombre de professionnels à formuler des critiques à l'encontre des modèles internes dynamiques du fait de leurs besoins de ressources pour la réalisation des calculs sur de nombreux tirages.

Toutefois, une question pourrait se poser sur ce degré de finesse des données en entrée des modèles internes. En effet, avoir à disposition des données agrégées est-il nécessairement un problème dans le cadre des modèles stochastiques à implémenter pour calculer les valeurs de portefeuille ? Des données agrégées de qualité ne permettent-elles pas de réaliser des calculs suffisants pour répondre aux normes *Market Consistent* ? Une évaluation réalisée sur un portefeuille dont le *business model* ne requiert pas de données individuelles fines, mais uniquement des données agrégées, doit-il être considéré comme non fiable ?

---
[1] MCEV : *Market Consistent Embedded Value*

De nombreuses études, comme celle de Juha M. Alho [2007] montrent que les tentatives faites pour incorporer des informations plus fines dans les projections n'ont pas forcément amélioré leur exactitude. Le présent article tente d'apporter une réponse pragmatique à la mesure du coût d'un type particulier d'options et garanties financières dans le cadre de la mesure du risque de marché, et à la mesure du risque de souscription vie dans une situation d'information incomplète, pour un portefeuille de garanties décès de prévoyance individuelle.

## 2. Risque de marché

Les principes directeurs de l'*Embedded Value* imposent d'utiliser une approche *Market Consistent* pour déterminer la valeur des actifs et des passifs en parfaite cohérence avec les marchés et entre eux. Ainsi, chaque flux doit être valorisé en utilisant le taux d'actualisation cohérent avec celui appliqué à des flux similaires sur les marchés de capitaux.

En pratique, il est possible de classer le traitement des flux suivant trois cas distincts. Pour les flux ne dépendant pas des mouvements de marchés, et ceux sans options et garanties intrinsèques, utilisation du taux sans risque par une approche en Equivalent Certain ; pour les flux dépendants des marchés avec options et garanties financières valorisables « à part », utilisation de l'approche en Equivalent Certain tout en valorisant les options intrinsèques par des techniques stochastiques ; pour les autres flux, utilisation de simulation de Monte-Carlo, en utilisant un modèle risque neutre, ou des déflateurs.

Dans cet article, nous supposerons que les flux de notre portefeuille ne dépendent pas des mouvements de marchés, hormis par l'existence d'une rémunération financière du distributeur sur le niveau de provisions affectées au portefeuille de contrats.

### 2.1. *La problématique*

Dans le cadre de l'étude des options et garanties financières sous-jacentes aux contrats, nous étudierons le cas de la rémunération financière du distributeur sur le niveau des provisions techniques. Ainsi, la situation la plus intéressante du point de vue actuariel est celui où le distributeur pourrait être rémunéré sur le niveau des provisions en lui reversant une commission calculée sur la base d'un indice de marché avec taux minimum garanti ; soit, plus simplement, le versement du rendement excédentaire de l'indice de marché par rapport au taux minimum contractuel est reversé au distributeur.

Cette politique de rémunération du distributeur peut s'apparenter à la politique de rémunération des contrats d'assurance vie comprenant un taux minimum garanti avec participation aux bénéfices.

### 2.2. *En termes financier*

Pour déterminer la valeur de marché d'un tel schéma, nous appliquons la méthode financière de réplication par des instruments financiers que sont les actifs de taux de marché.

Ainsi, si le taux de marché est supérieur au taux minimum, l'assureur payera un coupon d'excédent, et dans le cas contraire, aucun flux financier complémentaire ne sera versé [en sus du taux minimum contractuel].

En termes d'instruments financiers, un schéma de flux de ce type correspond au *payoff* d'un cap. Pour rappel, un cap est un contrat avec lequel le vendeur promet de rétribuer son porteur si le taux d'intérêt de référence vient à dépasser un niveau prédéterminé (appelé le taux d'exercice du cap) à certaines dates dans le futur.

Les caps sont des contrats liquides sur les marchés, et leur évaluation est basée sur la formule de « pricing » de Black [1976]. Ce modèle est une version du modèle BSM (Black et al. [1973]), adapté aux produits de taux d'intérêt.

Sur les marchés financiers, les caps sont composés de caplets qui sont côtés en volatilité issue de la formule de Black.

Par définition (cf. Hull [2005]), le *payoff* d'un caplet à la date $T_j$ s'écrit :

$$C_j = N.\delta.\max\left(0; R^L(T_{j-1}, \delta) - E\right)$$

Avec $N$ le nominal de l'opération, $R^L(T_{j-1}, \delta)$ le taux d'intérêt en $T_{j-1}$ de maturité $\delta$ mois, et $E$ le taux d'exercice du cap (ou encore le strike).

La variable diffusée dans le modèle de Black est le taux forward linéaire $F(t, T_{j-1}, T_j)$,

Dés lors, le payoff d'un caplet à la date $T_j$ peut s'écrire

$$C_j = N.\delta.\max\left(0; R^L(T_{j-1}, \delta) - E\right) = N.\delta.\max\left(0; F(T_{j-1}, T_{j-1}, T_j) - E\right)$$

La formule générale de valorisation d'un cap s'écrit alors

$$CAP(t) = \sum_{j=1}^{n} N * B(t, T_j) * \left[F(t, T_{j-1}, T_j) * \phi(d_j) - E * \phi\left(d_j - \sigma_j \sqrt{T_{j-1} - t}\right)\right]$$

Avec $F(t, T_j, T_{j-1})$, le taux forward calculé à la date t, $\sigma_j$ la volatilité du taux forward, et $\Phi$ la fonction de répartition de la loi normale centrée réduite de paramètre $d_j$ où

$$d_j = \frac{1}{\sigma_j * \sqrt{(T_{j-1} - t)}} * \left[Ln\left(\frac{F(t, T_{j-1}, T_j)}{E}\right) + \frac{\sigma_j^2}{2} * (T_{j-1} - t)\right].$$

Dans le cadre du modèle interne de projection, le montant nominal n'est pas fixe, il est variable en fonction du temps, et correspond au montant des provisions techniques du portefeuille étudié. En se plaçant à la date de calcul, la formule de valorisation du cap s'écrit

$$CAP(0) = \sum_{j=1}^{n} N_j * B(0, T_j) * \left[F(0, T_{j-1}, T_j) * \phi(d_j) - E * \phi\left(d_j - \sigma_j \sqrt{T_{j-1}}\right)\right]$$

Avec $N_j$ le montant de provisions techniques à la date *j*, et $E$ le taux minimum contractuel.

Finalement, la valeur temps de l'option avec garantie financière est la valeur du cap ainsi calculée.

### 2.3. Applications numériques

Nous supposerons l'existence d'une convention avec un distributeur avec les paramètres suivants de rémunération des provisions techniques.

- Taux minimum contractuel égal à 1,90 %,
- Taux de référence indexé sur le taux des emprunts d'états 3 ans,
- Volatilité implicite (maturité 3 ans) à la date t=0 égale à 16,60%.

Il est à noter qu'un paramètre doit être récupéré sur les marchés : la courbe de volatilité implicite des caps de maturité 3 ans.

La figure 1 présente la mesure du risque de marché lié à la rémunération des provisions techniques. Sur un niveau de provisions de 1,3 millions d'euros et pour les paramètres de rémunération définis ci-dessus, le coût du risque s'élève alors à 6 671 euros.

| Country | Country1 | | | | Partner Name | Partner1 | | | | Asset Rate | 3,41% |
|---|---|---|---|---|---|---|---|---|---|---|---|
| YearQuarter | 2005Q4 | | | | | | | | | Risk free rate | 2,26% |
| Volatility (at time 1) | 16,60% | | | | | | | | | 3 years rate | 1,95% |
| Technical Rate | 1,90% | | | | Taxes | 27,50% | | | | | |

| Time | 0 | 1 | 2 | 3 | 4 | 5 | 6 |
|---|---|---|---|---|---|---|---|
| Zero Coupon Rate Curve | 2,38% | 2,68% | 2,79% | 2,91% | 2,95% | 3,03% | 3,11% |
| Cash-flow | 1 327 916,50 € | 1 132 393,77 € | 578 046,58 € | 262 250,98 € | 95 451,83 € | 20 285,94 € | - € |
| Discounted Cash-flow | 1 327 916,50 € | 1 102 837,72 € | 547 044,19 € | 240 653,00 € | 84 964,97 € | 17 476,09 € | - € |
| Discount Factor | 1 | 0,973899494 | 0,946366958 | 0,917643843 | 0,890134519 | 0,861487903 | 0,832092274 |
| 3 years Forward Rate | 1,95% | 2,91% | 3,04% | 3,18% | 3,32% | 3,47% | 3,60% |
| Contractual Rate | 1,91% | 2,14% | 2,17% | 2,21% | 2,24% | 2,28% | 2,31% |
| Volatility | | 16,60% | 17,00% | 16,90% | 16,60% | 16,10% | 15,55% |
| Caps | 153,48 € | 3 175,22 € | 1 981,13 € | 1 019,51 € | 404,18 € | 91,44 € | - € |
| Projection | 153,48 € | 2 667,89 € | 1 503,45 € | 741,60 € | 289,22 € | 65,88 € | - € |

| Stochastic value | - 6 671,48 € |
|---|---|
| Deterministic value | - 5 268,03 € |
| Valorisation spread | 1 403,44 € |

| Cash-flow in P&L Account | - 638,46 € | 544,45 € | 277,92 € | 126,09 € | 45,89 € | 9,75 € | - € |
|---|---|---|---|---|---|---|---|
| Discounted Cash-flow | - 638,46 € | 530,24 € | 263,02 € | 115,70 € | 40,85 € | 8,40 € | - € |
| Sum of discounted cash-flow | - 958,21 € | | | | | | |

| CRO | - 4 142,12 € |
|---|---|

R&V : Cost of Remuneration Option

**Figure 1 : Mesure du risque de marché lié à la rémunération des provisions techniques**

### 3. Risque de souscription vie

Pour rappel, l'*Embedded Value* est une mesure de valeur composée de deux parties : le patrimoine accumulé au cours du temps et la valeur des contrats en portefeuille obtenue par projection en *run off* des flux de résultats générés.

Pour calculer la valeur du portefeuille, le principe général est de projeter le compte de résultat sur un horizon de temps donné (par exemple 30 ans), puis d'actualiser les flux de résultats.

En situation d'information incomplète, nous supposerons que le modèle est basé sur une projection des ratios de sinistralité actuariels, notés *S/P*, dont l'évolution déterministe est acquise à l'aide d'études actuarielles sur des portefeuilles similaires, suivant une table de mortalité par exemple (cf. Aguir [2006]).

Dans le cadre de la mesure spécifique du risque de souscription vie et de sa valorisation en normes MCEV, nous devons prendre en compte deux composantes importantes : la partie asymétrique du risque liée à un écart entre l'espérance et la médiane de la valeur présente des profits futurs, et la partie symétrique liée à une incertitude sur les flux futurs due à une aversion au risque.

Dans cet article, nous traiterons dans un premier temps la question de l'asymétrie, puis nous traiterons la question de la symétrie et de l'aversion au risque, et finalement nous proposerons une approche simplifiée pour les mesurer conjointement.

### 3.1. Risque asymétrique

#### 3.1.1. Commercialisation de produits sans intermédiaire

Dans le cas simplifié de la vente de produits sans intermédiaire (appelée vente directe dans le suite de l'article), le risque technique sous-jacent aux activités de prévoyance est symétrique et se limite à une éventuelle dérive du ratio de sinistralité, qui pourrait dans une modélisation déterministe, être évalué par un scénario central, issu d'une étude statistique du passé. Ce scénario est généralement une chronique future et espérée du ratio de sinistralité actuariel.

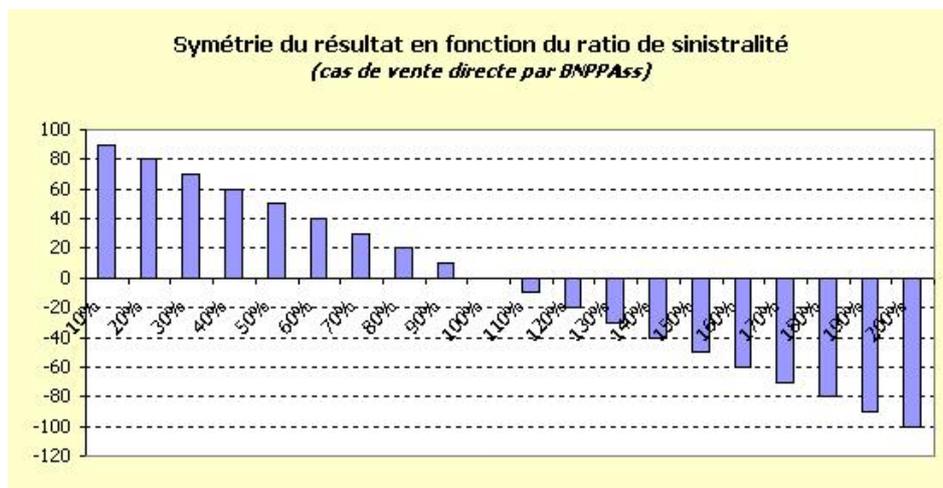

**Figure 2 : Cas de vente directe par l'assureur**

La figure 2 présente l'impact de la vente directe des produits sur le résultat de sinistralité assureur. Ainsi, nous supposons l'existence d'une prime de 100 UM[2], avec des ratios de sinistralité actuariel *S/P* variant de 10 % à 200 % et un taux de *profit sharing* de 0 % (vente directe donc pas de partage du résultat). Comme attendu, le résultat assureur est symétrique autour du ratio de sinistralité actuariel de 100 %.

#### 3.1.2. Commercialisation de produits avec intermédiaire

Dans le cas de la vente de produits par l'intermédiaire de partenaires ou de distributeurs, l'existence de clauses contractuelles de participation aux bénéfices avec ces partenaires ou distributeurs a pour conséquence de créer une asymétrie du résultat de l'assureur.

---

[2] UM = unités monétaires

Nous ne détaillerons pas le mécanisme de participation aux bénéfices ; néanmoins, nous pouvons en donner une vision simplificatrice[3] dans la figure ci-dessous.

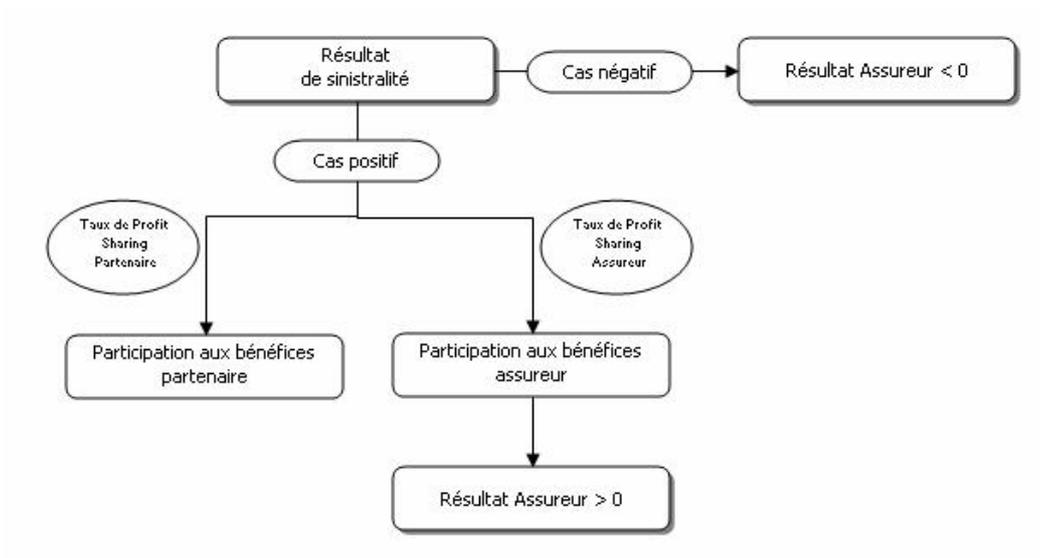

**Figure 3 : Mécanisme simplifié de participation aux bénéfices**

Le mécanisme est le suivant : lorsque le ratio de sinistralité actuariel est inférieur à 100 %, le résultat de sinistralité[4] est positif ; et dans le cas d'un partage des bénéfices, l'assureur versera la part contractuelle au distributeur, égal au taux contractuel appliqué au résultat de sinistralité.

Dans le cas contraire, lorsque le ratio de sinistralité actuariel est supérieur à 100 %, le résultat de sinistralité devient négatif, et dans le cas d'un partage des bénéfices avec le partenaire, ce résultat est alors intégralement supporté par l'assureur.

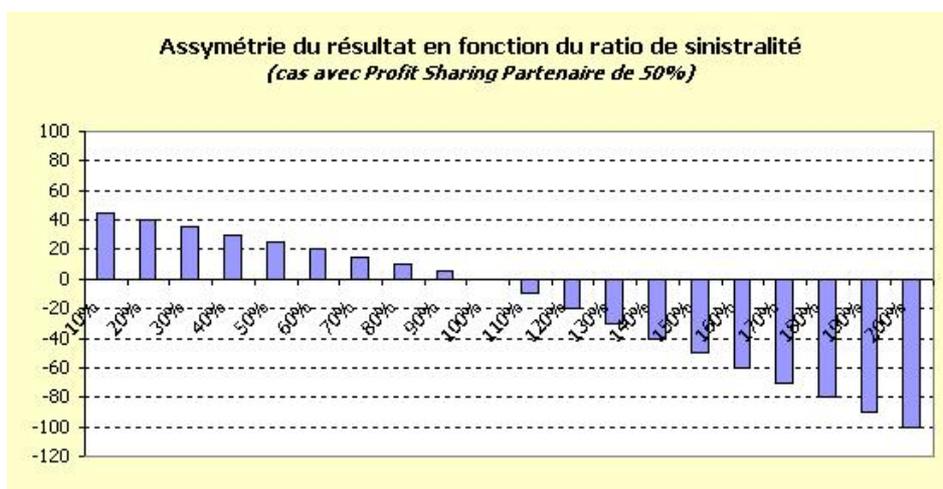

**Figure 4 : Cas de la vente par l'intermédiaire d'un distributeur**

---

[3] Le schéma ne tient pas compte de l'éventuelle existence des provisions d'égalisation ou réserves de stabilité
[4] Par soucis de simplification, nous supposerons : résultat de sinistralité = primes - sinistres

La figure précédente présente le cas de la vente de produits avec intermédiaire, i.e. un partenaire vendant les produits (cas relativement fréquent pour les compagnies d'assurance ayant des filiales à l'international). Ainsi, nous supposons l'existence d'une prime de 100 UM, avec des ratios de sinistralité actuariels variant de 10 % à 200 %, et un taux de *profit sharing* partenaire de 50 %.

Dans ce cas, nous pouvons remarquer un « tassement » du résultat pour les ratios de sinistralité actuariel *S/P* inférieur à 100 %, impliquant ainsi une asymétrie du résultat pour l'assureur, qui se caractérise par le fait que l'espérance du résultat conditionnellement au ratio de sinistralité, n'est pas égale à la valeur du résultat conditionnellement à l'espérance du ratio de sinistralité.

### 3.2. Idées sous-jacentes à la modélisation

La modélisation du risque de souscription est liée à la mise en place d'un grand nombre de scénarios de sinistralité futurs afin de tenir compte de la volatilité sous-jacente de la sinistralité à laquelle fait face l'assureur. Il s'avère alors nécessaire de mettre en place un modèle stochastique de projection de cette sinistralité.

Parmi les réflexions actuelles sur les modèles internes, la modélisation des actions du management nous amène à penser à l'idée d'un retour à la moyenne des ratios de sinistralité actuariels après un choc dans le cadre du *run-off* modélisé en normes MCEV. De manière simplificatrice, trois types de choc peuvent impacter les ratios de sinistralité :

- un choc structurel,
- un choc conjoncturel,
- un choc mixte conjoncturel et structurel.

Hormis le cas d'un choc conjoncturel où le ratio de sinistralité pourrait revenir à une évolution « normale », le modèle doit prendre en compte la possibilité que les actuaires puissent influencer l'évolution à court ou moyen terme l'évolution du ratio de sinistralité actuariel.

Ainsi, plusieurs exemples peuvent être donnés suivant le type de produits. Certains contrats vendus sont des contrats révisables annuellement, permettant ainsi à l'actuaire de revoir ses paramètres de tarification et de modifier sa prime de risque. Pour les contrats non révisables annuellement, l'actuaire peut avoir la possibilité de doter des provisions supplémentaires sous réserve du respect des Codes des Assurances : par exemple, par l'estimation de provisions mathématiques et/ou de provisions pour risques croissants (PPRC) complémentaires.

### 3.3. La méthodologie proposée pour modéliser l'évolution du ratio

Une première idée pour concilier l'idée d'un retour à la moyenne du ratio de sinistralité avec une modélisation stochastique, serait de retenir un modèle de taux de type Cox Ingersoll Ross [1985].

#### 3.3.1. Pourquoi ne pas retenir ce choix ?

Dans le cadre de l'article, ce choix n'a pas été retenu. Une des premières raisons concerne le fait que l'utilisation d'un modèle de taux nécessite d'effectuer une estimation fine du paramétrage comme par exemple, la moyenne à long terme. Ce point ne pouvant être respecté, du fait d'une absence d'informations concernant un éventuel historique à long terme du ratio de sinistralité - les ratios de sinistralité nets de tous les éléments (blanchiment, sur provisionnement, « reprise exceptionnelle » des provisions par exemple) n'étant pas forcément historisés sur tous les périmètres, produits ou pays.

La deuxième raison réside dans le risque de modèle qui est le principal inconvénient des scénarios stochastiques. En effet, si le modèle est mal calibré, tous les risques seront calculés avec l'erreur initiale, et il peut être dangereux d'utiliser les résultats obtenus pour la gestion et le calcul du coût des risques. Ce problème est d'autant plus important sur le long terme car les calculs sur un horizon long sont extrêmement sensibles aux hypothèses sous-jacentes du modèle. Si l'utilisateur utilise des distributions inappropriées ou des paramètres erronés dans le modèle, la mesure de risque peut aboutir à des résultats caducs qui peuvent se manifester sur l'horizon de projection.

Enfin, le dernier argument est la volonté de disposer d'un modèle ne nécessitant pas des simulations à chaque pas de temps de projection pour un horizon long sur certains périmètres dans l'idée de réduire le temps de calcul dans le cadre d'une valorisation « industrielle ».

### 3.3.2. *Méthodologie retenue*

Afin de pallier au manque de données et de simplifier le processus de calibrage des paramètres à utiliser, la méthode proposée est la suivante :

- Réaliser un tirage de *N* valeurs du ratio de sinistralité actuariel à la date « *t*=1 », suivant une loi à définir,
- Modéliser un retour à l'espérance des ratios de sinistralité actuariel avec paramètres simplifiés sachant le ratio de sinistralité en 1

#### 3.3.2.1 Première étape : en t=1, les tirages stochastiques

En *t*=1, par simulation, nous réalisons le tirage de *N* valeurs du ratio de sinistralité *S/P*(1) suivant une loi déterminée et de paramètres obtenus à partir des informations à notre disposition sur les portefeuilles valorisés.

L'hypothèse forte que nous retiendrons est de supposer que le ratio de sinistralité suit une loi Log Normale[5] de paramètres $\sigma$ et $\mu$. Par définition et propriétés de la loi Log Normale, cette loi LN ($\mu$, $\sigma$) a pour densité, espérance et variance

$$f(x) = \frac{1}{x\sigma\sqrt{2\pi}} e^{\left(\frac{-1}{2}\left(\frac{\log(x)-\mu}{\sigma}\right)^2\right)}, \; E(X) = e^{\mu+\frac{\sigma^2}{2}} \text{ et } V(X) = e^{2\mu+\sigma^2}(e^{\sigma^2}-1)$$

La variable X correspondant au ratio de sinistralité actuariel, nous obtenons après inversion des formules, les paramètres $\sigma$ et $\mu$ de la loi Log Normale.

$$\mu = Ln\left[E\left(S/P\right)\right] - \frac{\sigma^2}{2} \text{ et } \sigma = \sqrt{Ln\left(Vol\left(S/P\right)^2 + 1\right)}$$

---

[5] Entre autres raisons, la loi Log Normale est définie sur R+, et possède une queue plus épaisse que la loi Normale.

L'espérance et la volatilité du ratio de sinistralité, paramètres nécessaires aux tirages aléatoires sont ensuite obtenues à partir d'informations calibrées sur les portefeuilles valorisés.

Considérons tout d'abord que l'espérance du ratio de sinistralité est le ratio de sinistralité comptable à la date de calcul, ou bien un ratio de sinistralité estimé dans le cas où le ratio comptable aurait été sujet à retraitements exceptionnels.

Quant à la volatilité du ratio de sinistralité, nous supposerons que ce paramètre est expliqué par un critère qualitatif, l'ancienneté du portefeuille en années, et cinq critères quantitatifs, que sont l'homogénéité des capitaux sous risque, la qualité des bases techniques, le risque de concentration, le risque moral, et le risque de contentieux. Ces critères peuvent par exemple être obtenus à l'aide d'une analyse en composantes principales (ACP).

| Critères | Classification | | |
|---|---|---|---|
| Ancienneté du portefeuille | < 1 an | < 4 ans | > 4 ans |
| Homogénéité du portefeuille | Fort | Modéré | Faible |
| Qualité des bases techniques | Fort | Modéré | Faible |
| Concentration | Fort | Modéré | Faible |
| Risque moral | Fort | Modéré | Faible |
| Risque de contentieux | Fort | Modéré | Faible |

**Tableau 1: Qualification de la volatilité du ratio de sinistralité**

Le tableau ci-dessus fournit un exemple d'options possibles données aux modélisateurs pour caractériser « qualitativement » la volatilité du ratio de sinistralité des portefeuilles valorisés ; ces choix permettant alors d'allouer une volatilité numérique au ratio de sinistralité à partir d'un poids associé à chaque critère qualitatif ; la pondération associée à chaque cellule du tableau pouvant être déterminée par une régression statistique sur l'historique d'un portefeuille en données individuelles similaire.

Finalement, la volatilité du ratio de sinistralité est obtenue à partir de la pondération $\alpha_i$ allouée à chaque cellule du tableau ci-dessus.

$$Vol\left(S/P\right) = \prod_i \alpha_i$$

En tenant compte des remarques précédentes et de la log normalité supposée, la matrice des ratios de sinistralité actuariels, de dimension M($\omega$ x 1), s'écrit alors :

$$\forall i \in [1..\omega], \quad \frac{S}{P_i}(1) = \exp\left[\Phi^{-1}(u_i) \cdot \sigma_i + \mu_i\right]$$

Avec $\Phi^{-1}$ la fonction quantile de la loi normale réduite, et $u_i$ variables aléatoires uniformes définies sur [0,1], pour tout $i \in [1, \omega]$.

Afin d'illustrer le propos, nous proposons l'exemple suivant, basé sur les tirages de 10 000 ratios de sinistralité actuariels, en faisant varier le critère de qualité des bases techniques d'un risque faible (en rouge) à un risque fort (en bleu). Les densités sous-jacentes sont présentées dans les tableaux et figures ci-dessous.

Les hypothèses retenues pour les différents scénarios sont les suivantes :
- Espérance du ratio de sinistralité égale à 80%
- Volatilité du ratio de sinistralité basée sur des critères renseignés à risque modéré, sauf pour la qualité des bases techniques

| Critères | Seuil | Risque |
|---|---|---|
| Ancienneté du portefeuille | Moyen (3 ans) | Modéré |
| Homogénéité du portefeuille | Moyen | Modéré |
| Qualité des bases techniques | | |
| Concentration | Moyen | Modéré |
| Risque moral | Moyen | Modéré |
| Risque de contentieux | Moyen | Modéré |

**Tableau 2 : Critères retenus**

A partir des 10 000 tirages, nous présentons la densité suivant le niveau de risque considéré pour la qualité des bases techniques.

| S/P(1) | Qualité des bases techniques | | |
|---|---|---|---|
| | Risque fort | Risque modéré | Risque faible |
| [0%;10%[ | 0 | 0 | 0 |
| [10%;20%[ | 0 | 0 | 0 |
| [20%;30%[ | 32 | 1 | 0 |
| [30%;40%[ | 297 | 44 | 0 |
| [40%;50%[ | 783 | 358 | 41 |
| [50%;60%[ | 1312 | 1077 | 533 |
| [60%;70%[ | 1615 | 1791 | 1801 |
| [70%;80%[ | 1598 | 2091 | 2787 |
| [80%;90%[ | 1312 | 1721 | 2416 |
| [90%;100%[ | 947 | 1254 | 1476 |
| [100%;110%[ | 714 | 757 | 605 |
| [110%;120%[ | 523 | 436 | 244 |
| [120%;130%[ | 297 | 218 | 68 |
| [130%;140%[ | 186 | 119 | 22 |
| [140%;150%[ | 132 | 71 | 4 |
| [150%;160%[ | 99 | 34 | 1 |
| [160%;170%[ | 59 | 16 | 1 |
| [170%;180%[ | 34 | 6 | 0 |
| [180%;190%[ | 21 | 1 | 1 |
| [190%;200%[ | 10 | 1 | 0 |
| [200%;210%[ | 12 | 2 | 0 |
| [210%;220%[ | 9 | 0 | 0 |
| [220%;230%[ | 6 | 1 | 0 |
| [230%;240%[ | 0 | 0 | 0 |
| [240%;250%[ | 1 | 1 | 0 |
| [250%;260%[ | 1 | 0 | 0 |
| [260%;270%[ | 0 | 0 | 0 |
| [270%;280%[ | 1 | 0 | 0 |

**Tableau 3 : Déformation de la densité suivant le critère de qualité des bases techniques**

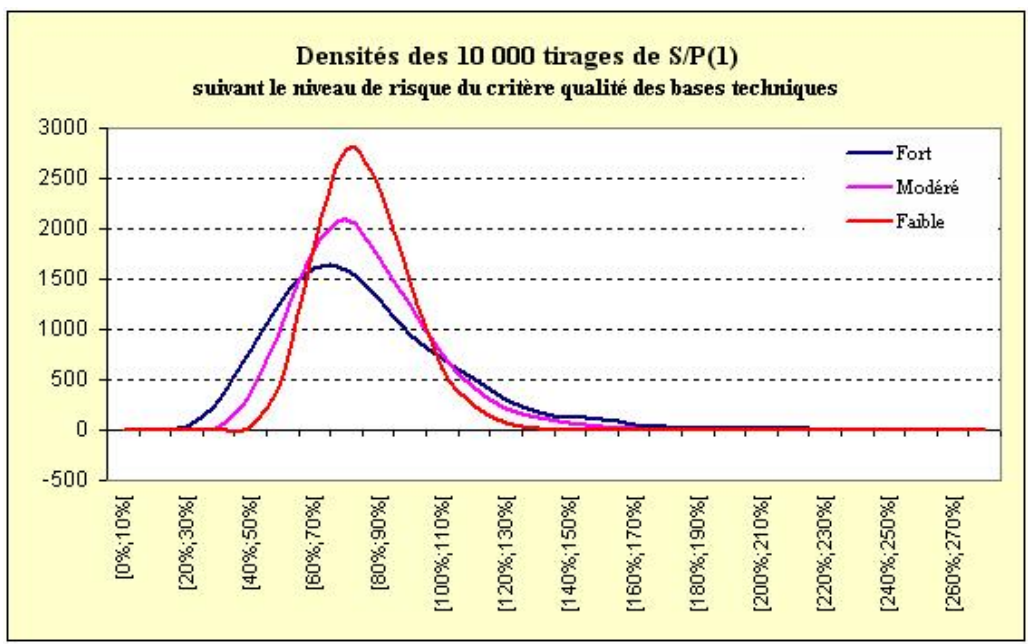

**Figure 5 : Déformation de la densité suivant le critère de qualité des bases techniques**

Comme attendu, l'analyse des résultats précédents montre que la déformation des densités se caractérise par des queues plus épaisses lorsque le risque augmente et plus fines lorsque le risque diminue. La figure précédente présente le scénario central en rose (c'est-à-dire que tous les critères sont renseignés comme risque modéré) ; comparativement, la courbe bleue associée à un risque fort sur la qualité des bases techniques s'aplatit et les queues de distribution s'épaississent ; quant à la courbe rouge associée à un risque faible, elle se recentre sur l'espérance (pour rappel, égale à 80%) et les queues de distribution se réduisent.

Sur la figure ci-dessous, un zoom sur l'épaisseur des queues de distribution à droite présente cette déformation, suivant le niveau de risque associé à la qualité des bases techniques.

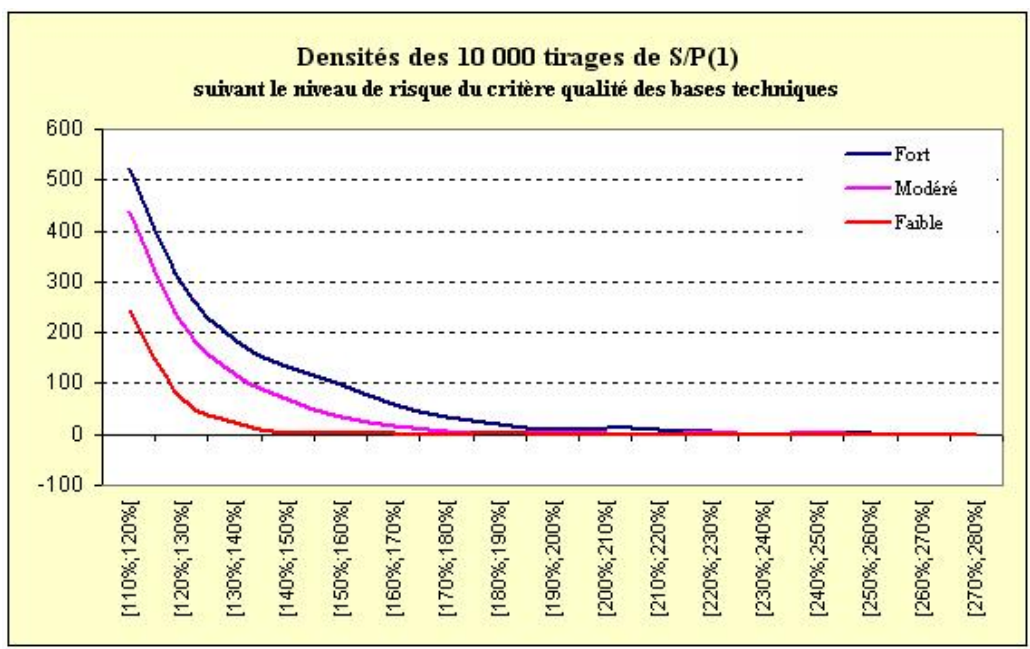

**Figure 6 : Déformation des queues de distribution**

### 3.3.2.2 Deuxième étape : modéliser le retour progressif à l'espérance

Comme précisé initialement, la méthode retenue consistera à effectuer une projection dynamique à partir de la chronique déterministe du ratio de sinistralité en tenant compte d'un retour simple et facilement paramétrable.

Finalement, $\forall t \in [2\,;30]$, pour chaque scénario de sinistralité, le ratio à la date *t* est obtenu par l'expression suivante avec *v*, la vitesse de retour à la moyenne.

$$\frac{S}{P_{i,j}}(t) = E\left[\frac{S}{P}(t)\right] + \left(\frac{S}{P_{i,j}}(1) - E\left[\frac{S}{P}(1)\right]\right) * v^{t-1}$$

Cette expression peut se décrire de la façon suivante, à l'espérance du ratio de sinistralité – c'est-à-dire l'évolution déterministe du ratio comptable à la date *t* -, nous ajoutons un facteur de dérive qui va assurer le retour progressif à la moyenne du ratio de sinistralité autour de la valeur déterministe avec une vitesse d'ajustement *v* déterminée.

### 3.3.2.3 Illustration de la méthode proposée

Pour illustrer le propos, dans les exemples présentés, nous supposerons que la vitesse de retour à la moyenne *v* est égale à 0,8. Cette valeur, basée sur l'étude d'un portefeuille représentatif, correspond à l'hypothèse qu'une durée de 3 ans est nécessaire pour réduire de moitié l'écart entre l'espérance et la valeur du ratio de sinistralité.

Etudions tout d'abord le cas du portefeuille 1 pour lequel l'âge actuariel est de 45 ans, le ratio de sinistralité à la date d'évaluation de 80 %, et avec pour critères de volatilité les éléments du tableau ci-dessous.

| Critères | Risque |
|---|---|
| Ancienneté du portefeuille | Modéré |
| Homogénéité du portefeuille | Faible |
| Qualité des bases techniques | Fort |
| Concentration | Fort |
| Risque moral | Faible |
| Risque de contentieux | Faible |

**Tableau 4 : Critères de volatilité du portefeuille 1**

Les critères précédents fournissent les paramètres de la loi Log Normale suivie par le ratio de sinistralité de ce segment : mu égal à -28 % et sigma égal à 33 %.

Pour simplifier, nous supposerons que le ratio de sinistralité est stable, et égale à la valeur retenue à la date de calcul, sur toute la durée de projection. Les figures ci-dessous (sur un horizon de 10 ans, puis de 30 ans) présentent la diffusion des scénarios stochastique qui est, comme nous l'attendions, centrée sur le scénario déterministe, et qui pour chaque ratio simulé présente un retour progressif à l'espérance.

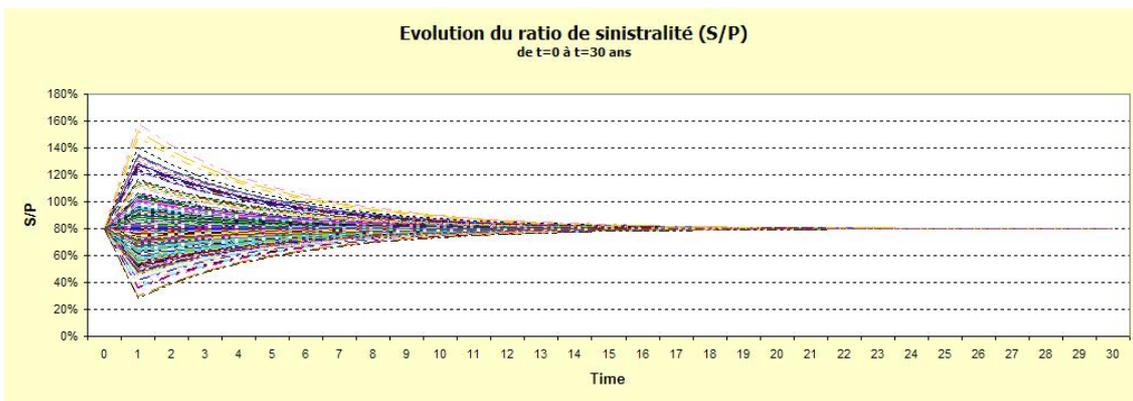

**Figure 7 : Diffusion des scénarios de sinistralité (horizon de 30 ans)**

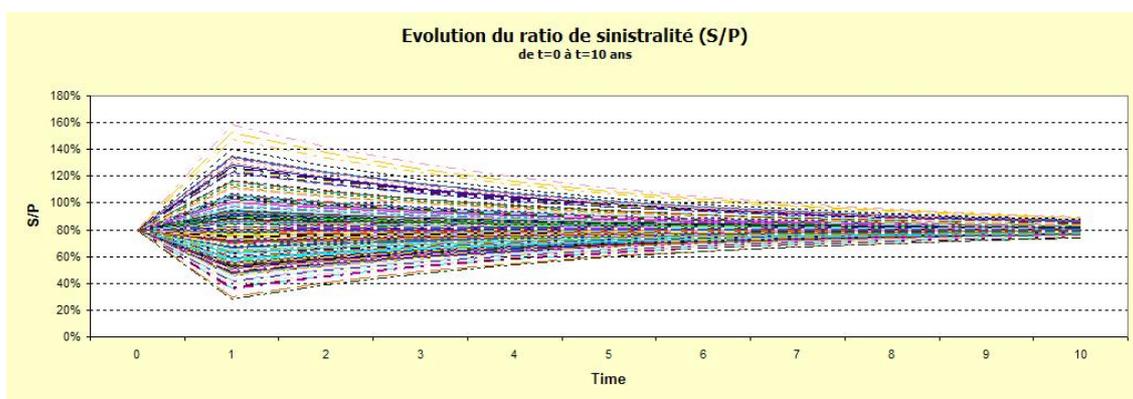

**Figure 8 : Diffusion des scénarios de sinistralité (horizon de 10 ans)**

Etudions maintenant le cas du portefeuille 2, pour lequel l'âge actuariel est de 45 ans, le ratio de sinistralité à la date de calcul de 104 % supposé croître sur un horizon de 10 ans, puis se stabiliser, et avec pour critères de volatilité les éléments du tableau ci-dessous.

| Critères | Risque |
|---|---|
| Ancienneté du portefeuille | Modéré |
| Homogénéité du portefeuille | Faible |
| Qualité des bases techniques | Fort |
| Concentration | Fort |
| Risque moral | Faible |
| Risque de contentieux | Faible |

**Tableau 5 : Critères de volatilité du portefeuille 2**

Les critères précédents fournissent les paramètres de la loi Log Normale suivie par le ratio de sinistralité de ce segment : mu égal à -2 % et sigma égal à 24 %.

Les figures ci-dessous (sur un horizon de 10 ans et de 30 ans) présentent la diffusion des scénarios stochastique qui est, comme nous l'attendions, centrée sur le scénario déterministe, avec la représentation du retour progressif à l'espérance des ratios.

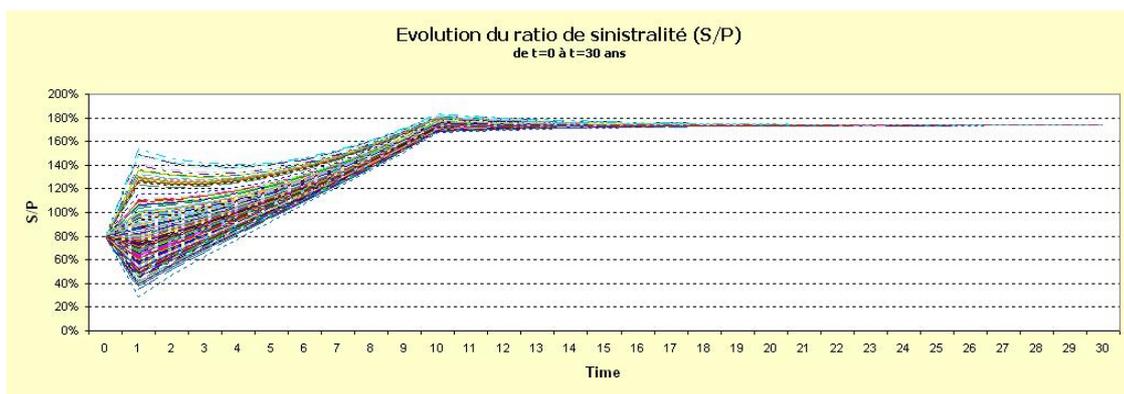

**Figure 9 : diffusion des scénarios de sinistralité sur un horizon de 30 ans**

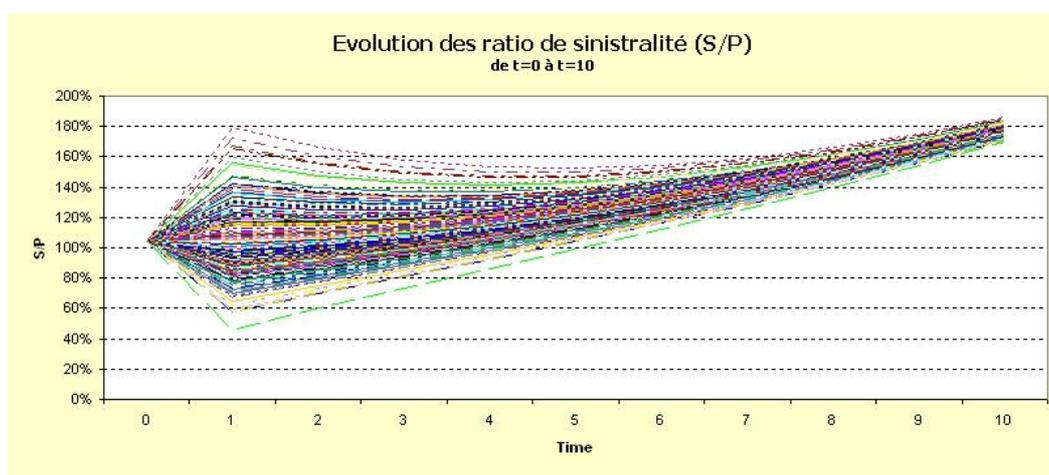

**Figure 10 : diffusion des scénarios de sinistralité sur un horizon de 10 ans**

### 3.4. Evaluation du coût du risque de souscription

En général, le coût du risque de souscription est donné par l'évaluation du coût des risques asymétriques liés à un écart entre la projection en scénario « central », c'est-à-dire le calcul de la valeur présente des profits futurs suivant un scénario d'évolution des ratios de sinistralité déterministe, et l'espérance de ce risque – soit l'écart entre l'espérance et la médiane en terme mathématique.

Toutefois, ne pas tenir compte de l'incertitude sur les flux futurs due à une aversion au risque peut s'avérer problématique. Un exemple concret pour représenter l'existence des risques symétriques est le cas des frais généraux : un coût de 100 UM dans un an impacte de 100 UM la valeur dans un an, alors qu'en considérant l'existence d'une aversion au risque de l'investisseur, cela aurait pu être un impact de 101 UM.

#### 3.4.1. Incertitude sur les flux futurs

En général, un investisseur accepte un certain niveau de risque pour un rendement donné. Ce niveau de risque est généralement représenté par la volatilité du résultat assureur.

Dans cet article, nous supposerons que ce niveau de risque est représenté par la volatilité de la valeur présente des profits futurs (PVFP) de notre portefeuille, elle-même liée à la volatilité du ratio, mais soumise à d'autres éléments financiers et opérationnels par exemple.

Ainsi, dans un premier temps, *N* comptes de résultat projetés sont mis en place à partir des *N* scenarii de cadences de *S/P* simulés. *N* valeurs présentes des profits futurs (notées PVFP(i)) (pour *i*= 1 à *N*) sont ensuite déterminées à partir de ces comptes de résultat projetés.

A partir de ces *N* valeurs, nous pouvons dès lors déterminer la moyenne des valeurs présentes des profits futurs, notée E[PVFP(TSR)], et la volatilité des valeurs, notée Vol[PVFP(TSR)].

De manière simplifiée, une volatilité élevée de la PVFP sera liée à une incertitude sur les flux futurs et entraîne donc une aversion de l'investisseur.

### 3.4.2. Tenir compte de l'aversion au risque de l'investisseur

#### 3.4.2.1 Couple rendement risque

La théorie financière de l'optimisation de la courbe rendement risque, ou *théorie moderne du portefeuille* développée en 1952 par Harry Markowitz (cf. Markowitz [1952]), nous apprend qu'il existe une relation positive entre le rendement et le risque. Cela revient à dire qu'un investisseur doit assumer plus de risque pour obtenir un rendement plus élevé.

#### 3.4.2.2 Représentation par un spread de taux

La méthode proposée par cet article sera de déterminer un *spread* lié à l'aversion au risque de l'investisseur, et basé sur la donnée initiale qu'est la volatilité d'un portefeuille d'assurés, i.e. le risque lié à ce portefeuille. En effet, chaque investisseur a son propre profil ; la plupart des investisseurs ont une aversion face au risque. Ils préfèrent un gain relativement sûr à un gain bien plus important mais aléatoire.

L'aversion au risque est l'un des tous premiers principes découverts en économie, principe issu d'un essai de « théorie sur la mesure du risque »[6] de Daniel Bernoulli (1738). Cette aversion au risque conduit au concept économique d'utilité et à la notion boursière de prime de risque, qui ont permis de mieux comprendre les équilibres de prix et de rendements, et d'aborder leur modélisation mathématique.

Les investisseurs ne seront prêts à prendre plus de risques qu'en échange d'un rendement plus élevé. A l'inverse, un investisseur qui souhaite améliorer la rentabilité de son portefeuille doit accepter de prendre plus de risques. L'équilibre risque rendement jugé optimal dépend de la tolérance au risque de chaque investisseur.

#### 3.4.2.3 Pour aller plus loin

Les recherches en finance comportementale ont montré que la notion d'aversion au risque doit être utilisée avec précaution ; en effet, les comportements des agents économiques vis-à-vis du risque et de l'espérance de gain étant en réalité très complexes et changeants.

Ainsi, la théorie conventionnelle ignore comment les « vraies » personnes prennent leurs décisions. Anas Tversky et Daniel Kahneman (cf. Kahneman et Tversky [1979]) ont été les pionniers dans ce domaine et ont remis en question l'hypothèse de rationalité absolue chez les investisseurs.

---

[6] Il s'agit d'un essai de mathématiques rédigé en latin dont le titre exact est « *Specimen theoriae novae de mensura sortis* ». L'essai porte sur l'équivalent entre une quantité certaine et une quantité hasardeuse (soit une variable aléatoire).

Ainsi, en 1979, Tversky et Kahneman ont défendu la position que des facteurs circonstanciels, tel que le contexte perçu de la décision, ont une influence prépondérante sur le comportement lié à la prise de risque.

Un très bon exemple à l'appui d'une vision de la prise de risque contingente à la situation dans laquelle elle est prise est donné par l'aversion individuelle au risque. D'un côté, on a montré que la plupart des individus sont réfractaires au risque dans des situations de gain, c'est-à-dire qu'ils préféreront un gain certain (300 euros) à une option risquée (80 chances sur 100 de gagner 400 euros), et ce, même si la valeur attendue de l'option risquée est supérieure (80% x 400 euros = 320 euros). D'un autre côté, les décideurs ont également tendance à être preneurs de risque dans des situations de perte, c'est-à-dire qu'ils préfèrent une option risquée (80 chances sur 100 de perdre 400 euros) à une perte certaine (300 euros).

La plus célèbre des théories proposées pour expliquer ces modifications du comportement face au risque dues à des différences de situation est la théorie des perspectives, selon laquelle les décideurs évaluent les options risquées à travers le prisme d'un système de valeurs subjectif, caractérisé par la dépendance à une référence et l'aversion aux pertes.

Pour l'essentiel, les individus sont censés coder les résultats relativement à un point de référence, tel que le statu quo, et vont ensuite les interpréter comme des pertes ou des gains. La théorie des perspectives suggère que les individus peuvent modifier leurs points de référence de telle sorte que leur comportement face au risque s'adapte en conséquence.

### 3.4.3. Lien retenu entre rendement et risque dans le modèle

Bien que les travaux présentés succinctement dans le paragraphe précédent permettent de mettre en évidence les impasses désormais reconnues de la théorie de l'utilité espérée, nous utiliserons cette théorie qui constitue toujours le paradigme dominant de la théorie de la décision en retenant un point de vue logique.

Ainsi en appliquant la théorie de la décision, il est acquis qu'un individu qui tolère moins le risque qu'un autre construit sa fonction d'utilité comme une fonction dont la concavité est plus prononcée.

#### 3.4.3.1 Lien retenu entre rendement et risque dans le modèle

La méthode proposée par cet article sera de lier les notions de rendement et de risque dans le modèle par le biais d'une fonction croissante et concave comme, par exemple, une fonction logarithme.

Ainsi, nous proposons de retenir la fonction suivante liant la volatilité du portefeuille (en rapport avec le risque) avec le spread (en rapport avec le rendement).

$$spread = f(\sigma) = a \ln(b\sigma + c)$$, avec a, b et c des constantes.

Afin de calibrer cette fonction logarithme, nous devons tenir compte du profil de l'investisseur et de son attente envers le marché.

Supposons que les activités de l'entité modélisées soient des activités d'épargne et de prévoyance, la fonction logarithme est alors supposée passer par, au minimum, les 3 points suivants :

- (spread prévoyance, volatilité prévoyance),
- (spread épargne, volatilité épargne),
- Origine (pas de risque, pas de spread).

Une analyse plus fine des activités ou des produits commercialisés pourrait être utilisée pour définir la fonction logarithme.

Pour illustrer le lien retenu, une estimation des constantes a, b et c, basée sur des données simplifiés – (2 %,10 %) et (3 %,20 %) – permettrait d'obtenir la fonction suivante :

$$spread = 0.0208 * \log(16.1878 * Vol[PVFP(TSR)] + 1)$$

A partir de la volatilité des valeurs de PVFP obtenues, nous pouvons en déduire le taux d'actualisation à appliquer aux flux de résultats pour tenir compte de l'incertitude en ajoutant le spread obtenu au taux sans risque utilisé dans le calcul en Equivalent Certain.

### 3.4.4. Evaluation du coût du risque

Comme nous l'avons montré précédemment, le coût du risque asymétrique de souscription vie est évalué par différence entre l'espérance et la médiane – c'est-à-dire le scénario déterministe -. L'existence d'une aversion au risque de l'investisseur sur le métier assurance sera intégrée comme un élément « réducteur » de cette espérance.

Le coût du risque de souscription vie s'écrit alors

$$CUR = PVFP_{TSR} - E[PVFP_{TSR}] * \frac{PVFP_{TSR+spread}}{PVFP_{TSR}}.$$

L'élément $\frac{PVFP_{TSR+spread}}{PVFP_{TSR}}$ caractérisant la prise en compte de l'aversion au risque de l'investisseur, en considérant que le même ratio peut-être appliqué à l'espérance.

### 3.4.5. Illustration de la méthode proposée

Dans le paragraphe suivant, nous présentons une illustration de la méthode proposée d'évaluation du risque de souscription vie.

#### 3.4.5.1 Les données initiales

La méthodologie proposée sera appliquée à trois portefeuilles de prévoyance individuelle dont les caractéristiques sont présentées dans les tableaux ci-dessous.

| Portefeuilles | Contrats renouvelables à tacite reconduction | Durée minimale des contrats à la souscription (en mois) | Durée maximale des contrats à la souscription (en mois) | Durée moyenne des contrats à la souscription (en mois) | Taux de chute | Stock de contrats |
|---|---|---|---|---|---|---|
| Portefeuille 1 | Non | 24 | 360 | 200 | | 34 261 |
| Portefeuille 2 | Oui | | | | 20% | 2 141 |
| Portefeuille 3 | Oui | | | | 20% | 15 505 |

**Tableau 6 : Données descriptives des portefeuilles valorisés**

| Portefeuilles | Ratio de sinistralité comptable | Ratio de sinistralité retenu | Age actuariel (en années) | Anticipation du risque |
|---|---|---|---|---|
| Portefeuille 1 | 46% | 95% | 55 | Oui |
| Portefeuille 2 | 13% | 40% | 44 | Non |
| Portefeuille 3 | 59% | 60% | 49 | Non |

**Tableau 7 : Données complémentaires nécessaires**

### 3.4.5.2 Les paramètres de la loi Log Normale

Les critères de volatilité de chacun des trois portefeuilles étudiés sont présentés dans le tableau ci-dessous.

| Portefeuilles | Ancienneté du portefeuille | Homogénéité des capitaux sous risque | Qualité des bases techniques | Risque de concentration | Risque moral | Risque de contentieux |
|---|---|---|---|---|---|---|
| Portefeuille 1 | 10 | Risque fort | Risque modéré | Risque faible | Risque modéré | Risque modéré |
| Portefeuille 2 | 9 | Risque fort | Risque modéré | Risque faible | Risque modéré | Risque modéré |
| Portefeuille 3 | 4 | Risque fort | Risque modéré | Risque faible | Risque modéré | Risque modéré |

**Tableau 8 : Critères de volatilité des portefeuilles valorisés**

Pour simplifier les calculs, seule différera le critère ancienneté du portefeuille, les autres données étant renseignées au niveau de la compagnie.

A partir de la volatilité du ratio de sinistralité déterminée et avec le ratio de sinistralité actuariel comptable ou modifié, nous évaluons les paramètres µ et σ de la loi Log Normale suivie par le ratio.

| Portefeuilles | Ratio de sinistralité comptable | Ratio de sinistralité retenu | Paramètre mu | Paramètre sigma |
|---|---|---|---|---|
| Portefeuille 1 | 46% | 95% | -7% | 19% |
| Portefeuille 2 | 13% | 40% | -94% | 21% |
| Portefeuille 3 | 59% | 60% | -54% | 26% |

**Tableau 9 : Paramètres de la loi Log Normale pour les portefeuilles valorisés**

### 3.4.5.3 Représentation graphique de trois portefeuilles

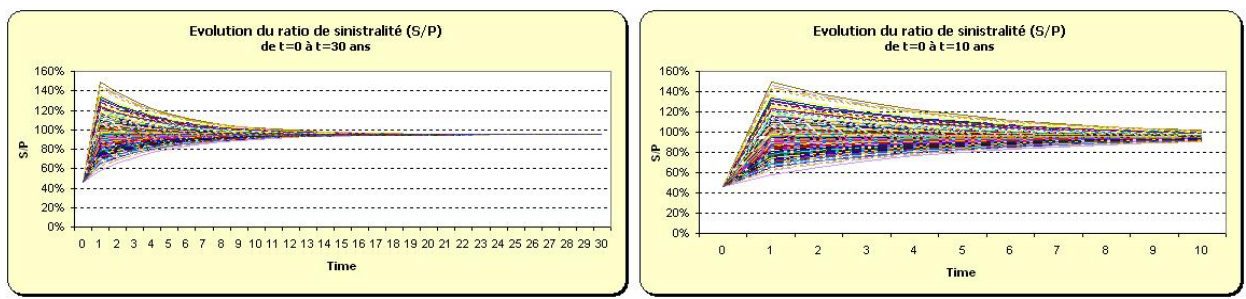

**Figure 11 : Diffusion des scénarios de sinistralité du 1er portefeuille**

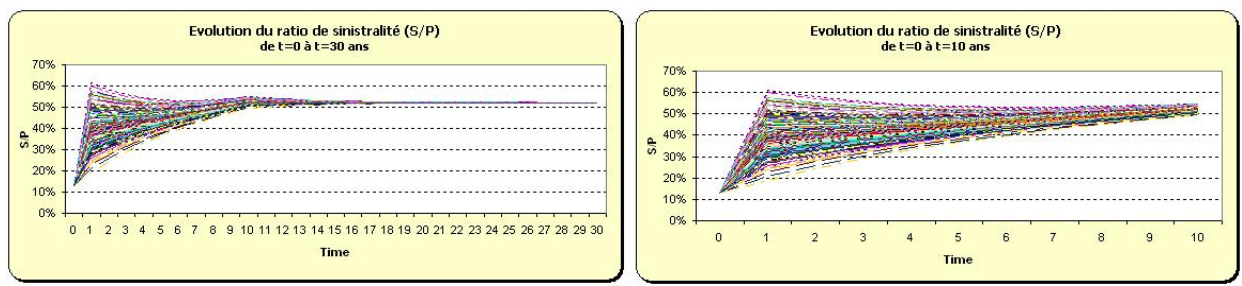

**Figure 12 : Diffusion des scénarios de sinistralité du 2ème portefeuille**

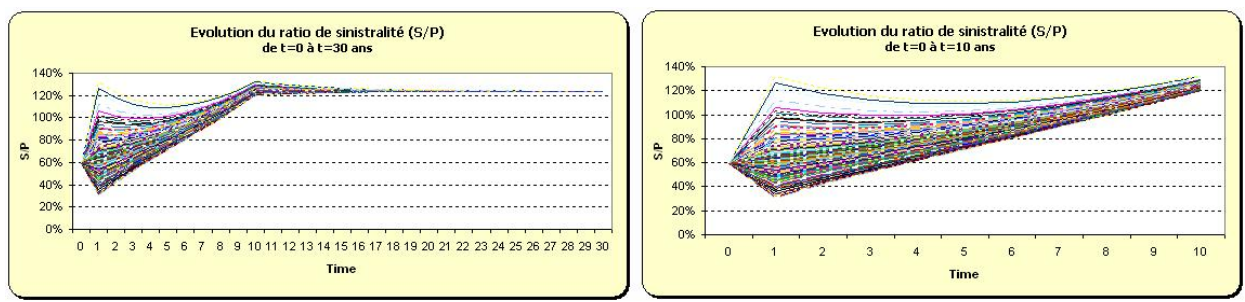

**Figure 13 : Diffusion des scénarios de sinistralité du 3ème portefeuille**

### 3.4.5.4 Les résultats

Les scénarii de ratios de sinistralité générés sont ensuite utilisés pour fournir l'espérance de la PVFP égale à la moyenne des *N* PVFP calculées, la volatilité de la PVFP égale à la volatilité des *N* PVFP calculées.

Le spread sera déterminé à partir de la fonction *spread* volatilité, et enfin nous obtenons la PVFP calculée avec un taux d'actualisation égal au TSR majoré du *spread*.

| Portefeuilles | Espérance de la PVFP | Volatilité de la PVFP | Spread de risque | PVFP (TSR + Spread) | PVFP (TSR) |
|---|---|---|---|---|---|
| Portefeuille 1 | 1 150 605 | 3 691 | 0,11% | 1 147 283 | 1 150 685 |
| Portefeuille 2 | 54 674 | 1 074 | 0,57% | 53 265 | 54 674 |
| Portefeuille 3 | 855 937 | 48 523 | 1,35% | 805 540 | 853 469 |
| Total | | | | | 2 058 828 |

**Tableau 10 : Eléments constitutifs du coût du risque de souscription**

| Portefeuilles | Espérance de la PVFP |
|---|---|
| Portefeuille 1 | 3 482 |
| Portefeuille 2 | 1 409 |
| Portefeuille 3 | 45 600 |
| Total | 50 491 |

**Tableau 11 : Evaluation du coût du risque de souscription**

## 4. Conclusion

Dans notre article, nous avons tenté d'apporter une réponse pragmatique aux questions posées en introduction sur le degré de finesse des données en entrée des modèles internes. Pour cela, à partir de portefeuilles de prévoyance individuelle dont le business model ne requiert pas de données individuelles fines, nous avons proposé une modélisation de la mesure du coût d'un type particulier d'options et garanties financières dans le cadre de la mesure du risque de marché et une modélisation de la mesure du risque de souscription vie.

Dans un premier temps, par application de la méthode financière de réplication par des instruments financiers, nous avons déterminé la valeur de marché de l'option liée à la rémunération financière, indexée sur un indice de marché, d'un distributeur.

Ensuite, afin de mesurer le coût du risque de souscription vie dans le cadre des risques non couvrables, nous avons proposé une approche permettant de mesurer les deux composantes du risque que sont la partie asymétrique, liée à un écart entre l'espérance et la médiane, et la partie symétrique, liée à une incertitude sur les flux futurs.

Concernant la partie asymétrique, la méthodologie proposée d'un calcul en espérance conditionnelle par rapport au ratio de sinistralité de l'année 1 permet de pallier au manque de données en utilisant la connaissance que les actuaires ont acquis sur ces portefeuilles ; enfin, cette méthodologie permet de réaliser les simulations stochastiques nécessaires à la prise en compte de la volatilité de la sinistralité, à laquelle fait face l'assureur. Pour la partie symétrique du risque, la méthodologie proposée utilise un nombre limité de données liées au profil de l'investisseur et de son attente envers le marché, ou les produits.

Comme pour tous modèles, la qualité des données est primordiale que nous soyons en possession de données individuelles fines ou de données fortement agrégées. Toutefois, les méthodes proposées, utilisables dans des situations d'information incomplète, répondent aux critères *Market Consistent* des principes directeurs de l'*Embedded Value*.

**Références**

# SOMMAIRE